\documentclass[twocolumn,showpacs, aps, superscriptaddress, eqsecnum, prx, notitlepage, showkeys, nofootinbib, floatfix]{revtex4-2}

\usepackage{graphicx}
\usepackage{dcolumn}
\usepackage{bm}
\usepackage[english]{babel}

\usepackage{amssymb}
\usepackage{amsmath}
\usepackage{graphicx}
\usepackage{dcolumn}
\usepackage{hyperref}
\usepackage{color,units}
\usepackage[dvipsnames]{xcolor} 
\usepackage{orcidlink}

\usepackage[acronym]{glossaries} 
\newacronym{HR}{HR}{high-reflective}
\newacronym{AR}{AR}{anti-reflective}
\newacronym{ppm}{ppm}{parts-per-million}
\newacronym{ITM}{ITM}{input test mass}
\newacronym{ETM}{ETM}{end test mass}
\newacronym{aLIGO}{aLIGO}{Advanced LIGO}
\newacronym{OPD}{OPD}{optical path depth}
\newacronym{TCS}{TCS}{thermal compensation system}

\definecolor{red}{rgb}{0.9, 0.1, 0.1}

\hypersetup{colorlinks=true, citecolor=teal, urlcolor=teal, linkcolor=teal, breaklinks=true}
\usepackage[commentmarkup=uwave,commandnameprefix=ifneeded]{changes}
\definechangesauthor[name=Max, color=teal]{MI}

\begin{document}

\title{You only thermoelastically deform once: \\
Point Absorber Detection in LIGO Test Masses with YOLO}

\author{\orcidlink{0000-0002-9575-5152} Simon R. Goode}
\email{simon.goode@monash.edu}
\affiliation{School of Physics and Astronomy, Monash University, VIC 3800, Australia}
\affiliation{OzGrav: The ARC Centre of Excellence for Gravitational Wave Discovery, Clayton, VIC 3800, Australia}

\author{\orcidlink{0000-0001-9298-004X} Mitchell Schiworski}
\affiliation{OzGrav: The ARC Centre of Excellence for Gravitational Wave Discovery, Adelaide, SA 5005, Australia}
\affiliation{School of Physics, Chemistry and Earth Sciences, University of Adelaide, Adelaide, SA 5005, Australia}
\affiliation{College of Arts and Sciences, Syracuse University, Syracuse, New York 13244, USA}

\author{\orcidlink{0000-0001-7851-3939} Daniel Brown}
\affiliation{OzGrav: The ARC Centre of Excellence for Gravitational Wave Discovery, Adelaide, SA 5005, Australia}
\affiliation{School of Physics, Chemistry and Earth Sciences, University of Adelaide, Adelaide, SA 5005, Australia}

\author{\orcidlink{0000-0002-4418-3895} Eric Thrane}
\affiliation{School of Physics and Astronomy, Monash University, VIC 3800, Australia}
\affiliation{OzGrav: The ARC Centre of Excellence for Gravitational Wave Discovery, Clayton, VIC 3800, Australia}

\author{\orcidlink{0000-0003-3763-1386} Paul D. Lasky}
\affiliation{School of Physics and Astronomy, Monash University, VIC 3800, Australia}
\affiliation{OzGrav: The ARC Centre of Excellence for Gravitational Wave Discovery, Clayton, VIC 3800, Australia}

\begin{abstract}
{Current and future gravitational-wave observatories rely on large-scale, precision interferometers to detect the gravitational-wave signals. 
However, microscopic imperfections on the test masses, known as \textit{point absorbers}, cause problematic heating of the optic via absorption of the high-power laser beam, which results in diminished sensitivity, lock loss, or even permanent damage. 
Consistent monitoring of the test masses is crucial for detecting, characterizing, and ultimately removing point absorbers.
We present a machine-learning algorithm for detecting point absorbers based on the object-detection algorithm \texttt{You Only Look Once (YOLO)}.
The algorithm can perform this task in situ while the detector is in operation.
We validate our algorithm by comparing it with past reports of point absorbers identified by humans at LIGO. 
The algorithm confidently identifies the same point absorbers as humans with minimal false positives.
It also identifies some point absorbers previously not identified by humans, which we confirm with human follow-up. 
We highlight the potential of machine learning in commissioning efforts.}
\end{abstract}

\maketitle
\section{Introduction}\label{sec:introduction}
Current and future gravitational-wave observatories rely on large-scale laser interferometers to observe gravitational-wave signals. 
To enhance the sensitivity of gravitational-wave observatories, researchers aim to increase circulating laser power to reduce photon shot noise. 
However, increasing the laser power presents a swathe of instrumentation challenges that must be overcome~\cite{Goodwin2024, Vinet2009}. 
One challenge is the absorption of the laser power, which causes thermal deformations in the interferometer mirrors that result in excess light scattering~\cite{Strain_1994}.

The high laser power in the arms of these interferometers is achieved via a pair of high-quality Fabry-Perot optical cavities, referred to as the power-recycling and arm cavities.
The mirrors that make up these arm cavities are called the \textit{test masses}.
These cavities require extremely low optical loss from absorption and scattering to maximise the optical gain in the cavities.
In this work, we are only concerned with scattering from thermally induced aberrations in the test masses.
Minor imperfections in the mirror coatings, known as \textit{point absorbers}~\cite{Brooks2021}, absorb excess laser power.
Concentrated heating induces high spatial frequency deformations on the test mass surfaces and their substrates. Such thermal defects scatter light out of the fundamental cavity mode, increasing optical loss and reducing detector performance.
However, the potential scattering into high-order cavity modes, which subsequently co-resonate, is also detrimental to the detector operation in various complex ways~\citep{Jia2021}, which leads to excessive commissioning time:

\begin{itemize}
    \item Degradation of the control systems required to maintain detector operation
    \item Positioning the beams off-centre, leading to increased coupling of radiation pressure instabilities~\citep{Evans2015}
    \item Wavefront mismatching of the beams in each arm
\end{itemize}

In the second observing run, the performance of both LIGO detectors was hindered by point absorbers~\citep{Buikema2020}, and they remain present in the fourth observing run~\cite{Cahillane2022} (Capote et al., in preparation)
The cause, prevention, and quantification of these point absorbers is an active area of research. As the optical power increases in future detectors, weaker point absorbers will become a problem once again.

This paper describes the development of a machine-learning algorithm designed to automatically detect point absorbers in the LIGO detectors. 
The algorithm utilises a deep convolutional neural network \citep{Lecun2015}, leveraging the pre-trained \textsc{You Only Look Once} (\texttt{YOLO}) framework \citep{Redmon2015} and transfer learning.
The algorithm's inputs are measurements of the test mass thermal deformation from the existing Hartmann wavefront sensors~\citep{Brooks2007,Brooks2016}.

We structure the paper as follows. 
Section \S\ref{sec:gw-detectors} introduces thermal deformation, point absorbers, and the detection of point absorbers with Hartmann wavefront sensors. 
Section \S\ref{sec:methods} describes how we generate synthetic data, simulate point absorbers on LIGO test masses at thermal equilibrium, and apply transfer learning with this data into the \texttt{YOLO} framework. 
Section \S\ref{sec:results} presents results from archival reports of point absorbers on LIGO test masses (available through publicly available LIGO ``alogs'') and compares them to model predictions. 
Finally, section~\S\ref{sec:conclusions} summarises the capabilities of the transfer-learned \texttt{YOLO} model, identifies its shortcomings, and highlights future work and improvements.

\section{Thermal aberrations}\label{sec:gw-detectors}

\subsection{Overview}

Absorption of the laser beam occurs in all optics within the detector.
This results in a loss of optical power, which is detrimental to the detector's sensitivity.
However, when the absorbed power is significant enough, one must also consider the heating of the optic and resultant thermo-optic effects.
Extensive research and development into glass production and optical coating techniques have allowed for extremely high-quality optics that absorb $\lesssim 1$~parts-per-million (ppm) of the incident laser power~\citep{Steinlechner2018,Granata2020}.
Still, absorption effects remain significant for the optics that experience the most laser power: the input test masses (ITMs) and end test masses (ETMs), which comprise the Fabry-Perot cavities in each arm of the detector.

Currently, the Advanced LIGO (aLIGO) detectors are commissioned to operate with $\unit[\sim400]{kW}$ of laser power circulating within each arm cavity~\citep{Cahillane2022}.
At design sensitivity, this number will increase to $\unit[750]{kW}$~\citep{Collaboration2015}, and for future generation detectors it will surpass $\unit[1]{MW}$~\citep{Hall2022}.
The inner-facing high-reflective coated surfaces of the test masses experience this optical power.
For these coatings, the absorption level is nominally $\unit[0.5]{ppm}$, meaning around half a Watt of heat is deposited into the optics.
Uncompensated, this seemingly small heat deposition is enough to affect the stability and sensitivity of the detectors.
The temperature of the optic increases typically by no more than a few degrees Celsius~\citep{Brooks2016}.
However, the heating is uneven, leading to a temperature change with a spatial distribution that causes optical aberrations in the resonating laser beam.

For LIGO's fused-silica test masses, the most impactful aberrations occur primarily from thermo-refractive lensing in the substrate, whereas thermal expansion is a smaller effect due to fused-silica's low thermal expansion coefficient.
Thermal lensing is the associated change in the refractive index with changing temperature, and the resulting aberrations are the most severe for a beam transmitted through the optic, such as the ITM.
Thermal expansion creates bulges on the surfaces of the optic, which are most severe when a beam is reflected from that surface~\citep{Lawrence2003}.
The impact of these thermal aberrations on the interferometer can be separated into those with spatially uniform absorption and those with non-uniform absorption of the laser beam.

Spatially uniform absorption of laser power within the coatings and its impact on the detectors has been extensively studied~\citep{Winkler1991,Lawrence2003,Vinet2009}.
In this scenario, the resultant heating profile matches the Gaussian intensity profile of the incident high-power laser beam.
For a beam that is transmitted through the optic, the resultant distortions induce, to the first order, a quadratic wavefront change (lensing effect).
For a beam that is reflected off the optic, the first-order effect is an apparent change in the radius of curvature.
These both may be counteracted in aLIGO by the thermal compensation system (TCS)~\citep{Brooks2016}, which applies additional heating to minimise the temperature gradient, thereby reducing optical distortions.

In practice, however, power is not absorbed uniformly.
Localised point absorbers~\citep{Brooks2021} appear on optics due to imperfections in the manufacturing process or micro-damage from vacuum particulates that sometimes land on the test mass.
These typically sub-micron-sized contaminants have absorption values that are orders of magnitude above nominal coating absorption.
Point absorbers present an experimental challenge as they are a source of loss and aberration that cannot be ameliorated with the thermal compensation system due to limited spatial resolution of actuators.
Therefore, it is essential to monitor the test masses in situ during the running of the detector to determine where they are on the test masses, how impactful they are, and if any new ones develop or previous ones change over time.

\subsection{Hartmann wavefront sensors}
\begin{figure}
    \centering
    \includegraphics[width=0.9\linewidth]{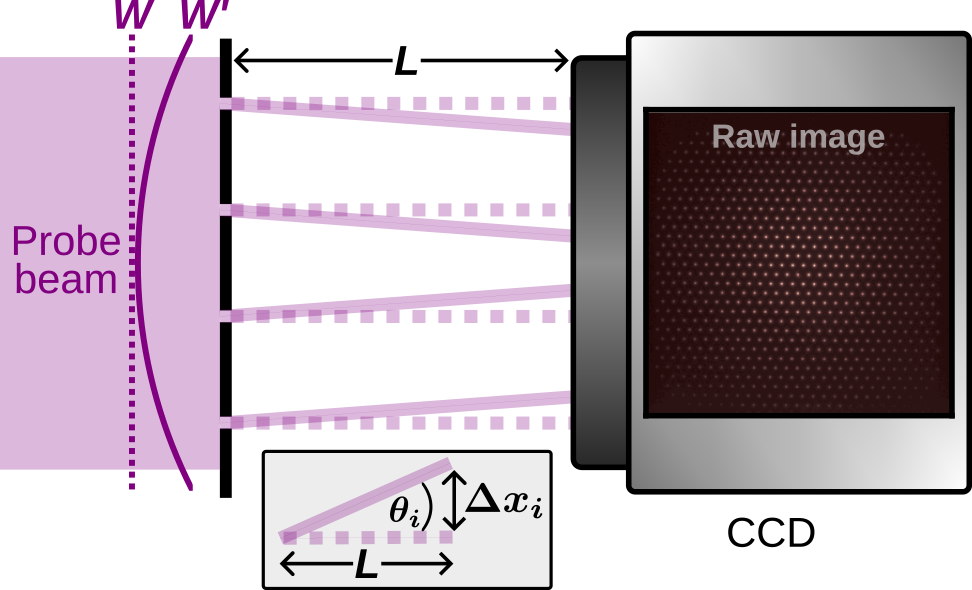}
    \caption{
    Diagram of a Hartmann wavefront sensor. 
    The probe beam is incident on a plate with a regularly spaced grid of holes located at a distance $L$ from a CCD sensor.
    Light diffracts through the holes and onto the sensor, producing an image of corresponding \textit{centroids} as shown in the example raw image.
    The angle that the light ray is diffracted, $\theta_i$ (and hence the position of the produced centroid) is equal to the gradient of the incident wavefront at the hole.
    A perfectly planar wavefront $W$ is not diffracted and produces centroids on the CCD aligned with the apertures, as indicated by the dashed lines.
    Some non-planar wavefront, e.g., $W'$, is diffracted at each hole by some angle, producing centroids displaced by some distance $\Delta x_i$.
    The gradient of the probe beam's wavefront, $dW'/dx$ is given by $dW'/dx = \theta_i \approx \Delta x_i/L$.
    The reference wavefront, in general, does not need to be planar.
    In this context, we use a reference image taken in the low laser power interferometer state so that $W'$ contains only information about the thermally induced deformations in the test mass.
    }
    \label{fig:HWS}
\end{figure}

The thermal deformation of a test mass is determined by measuring the shape of a wavefront reflected from its surface with Hartmann wavefront sensors.
These are components of the already existing TCS~\cite{Brooks2016}.
A Hartmann wavefront sensor, seen in Fig.~\ref{fig:HWS}, uses a camera (CCD or CMOS) and a Hartmann plate containing a grid of regular small apertures.
Perfectly planar wavefronts arriving parallel to the aperture plate pass through and present a grid of centroids on the sensor aligned with the aperture plate. 
Wavefronts that are not planar/parallel create a grid of centroids that deviate from the parallel planar alignment. 
One may infer the shape of the incoming wavefront by measuring how each centroid deviates from its reference point.

\begin{figure}
    \centering
    \includegraphics[width=0.9\linewidth]{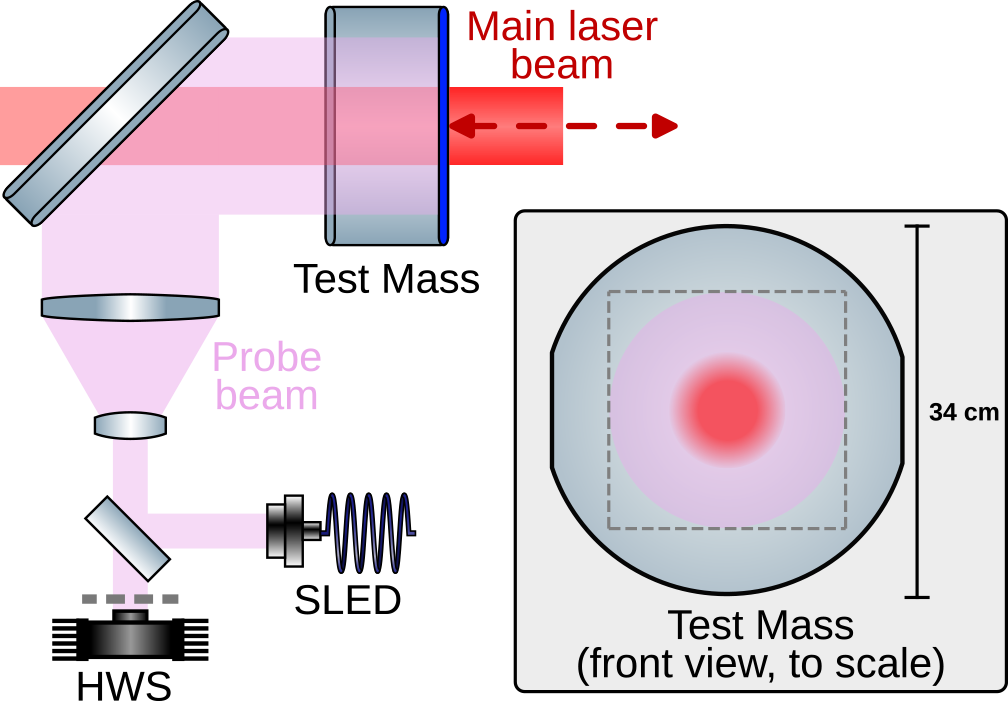}
    \caption{
    Simplified schematic of the scheme employed to monitor deformations in the test masses using Hartmann wavefront sensors.
    The probe beam is magnified and projected onto the test mass and retro-reflects the front coating onto the Hartmann wavefront sensor (HWS).
    The relative sizes of the main laser beam (red) responsible for the heating and the probe beam (purple) are shown.
    The dashed line corresponds to the actual viewing window of the test mass in the images produced.
    }
    \label{fig:HWS_telescope}
\end{figure}

Within aLIGO, a Hartmann wavefront sensor system is installed to monitor each of the four test masses as shown in Fig.~\ref{fig:HWS_telescope}. 
Each system has a superluminescent diode (SLED) probe beam that shines onto the test mass through a periscope and magnifying telescope. 
The probe beam is reflected from the test mass into a Hartmann wavefront sensor.
During operation, a reference set of Hartmann centroids is taken at the start of powering up the detectors (input laser is approximately 2W) in the so-called ``cold state''.
The cold-state measurement captures the curvature of the test mass with no thermal lensing and serves as a reference to measure thermally induced changes to the centroids when the test mass is heated.
The wavefront deformation of the probe beam $W'(x,y)$ in this scenario is then mostly:
\begin{equation}
    W'(x,y) \approx \int\frac{dn}{dT}\Delta T (x,y,z) \, dz
\end{equation}
where the scalar $dn/dT$ is the material thermo-refractive constant, and $\Delta T(x,y,z)$ is the temperature change of the test mass. The $z$ direction is the axis on which the probe beam propagates.
There are also smaller contributions from the displacements of the front and rear surfaces of the test mass, which occur via thermal expansion. 
This can then be interpreted as effectively measuring the temperature change of the test mass integrated through the optic.

A dedicated algorithm (external to this work) takes the two sets of Hartmann centroids, identifies their location in each image and measures their displacement.
It is then scaled to create a calibrated measurement of $\nabla W'(x,y)$, which is the gradient of optical path deformation of the probe beam.
In this format, the vector field $\nabla W'(x,y)$ is difficult to interpret by eye, therefore experts often perform a two-dimensional spline integration of this to estimate the optical path depth $W'(x,y)$. Although this is more intuitive, the calculation introduces interpolation/integration errors and can be computationally expensive. For these reasons, in this work, the algorithm is trained on the gradient field $\nabla W'(x,y)$, however we also include $W'(x,y)$ for comparison in our results.

Manually inspecting Hartmann data to identify point absorbers has some drawbacks:
\begin{itemize}
    \item It can only be carried out by a small number of trained experts whose time is often needed on other tasks.
    \item The results are somewhat subjective: while some point absorbers are easily identified, other candidates are not uniformly labelled by all experts.
    \item There is a limited dataset of point source measurements.
\end{itemize}
These limitations prohibit large-scale analyses of point absorbers, for example, investigating how point absorbers evolve with time as the test mass heats up, or any long-term variations in power being absorbed. With this work, our automated method can track point absorbers over long timescales with minimal human intervention.

\section{Method}\label{sec:methods}
Identifying point absorbers in Hartmann data is, at its core, a pattern recognition problem, which lends itself to a machine-learning solution.
Convolutional neural networks have been previously applied to improve the analysis of Hartmann-based sensors. This has often been applied for finding efficient wavefront integration methods for converting the Hartmann spot patterns into a measured wavefront or Zernike decomposition~\cite{Hu_Hu_Zhang_Gong_Si_2020, He_Liu_2021, Gu_Zhao_2021}. In contrast, our work is focused on object detection within Hartmann data.
\texttt{You Only Look Once} (\texttt{YOLO}; version \texttt{v8-n}), is a deep convolutional neural network algorithm created by Ultralytics\footnote{\url{https://docs.ultralytics.com/}}, which has been pre-trained to find a large variety of objects in images.
The choice to use a pre-trained architecture allows us to take advantage of \textit{transfer learning}. 
Transfer learning preserves a model's pre-trained weights and biases, particularly the components of the architecture responsible for feature extraction, but it allows the final layer(s) to adjust.
This allows us to leverage the extensive pattern recognition capabilities of the existing \texttt{YOLO v8-n} network and train it to detect point absorbers in the gradient field.
The following subsections describe the process for generating synthetic data and training \texttt{YOLO} to recognise point absorbers.

\subsection{Data Generation}\label{sec:methods-datagen}
To train the machine learning model, we require a large set of labelled data. 
Each sample of the dataset consists of a gradient field. 
The label for each sample is a bounding box centred on each point absorber and beam spot. 
We generate a synthetic dataset to ensure our training data contains a diverse range of point absorbers with known locations.
Generating the synthetic dataset consists of five steps, explained in detail in the following subsections:
\begin{enumerate}
    \item Generate synthetic optical path depths $W'(x,y)$ for test masses at thermal equilibrium with a random number of point absorbers of randomised absorption levels.
    \item Sample the average gradients of the optical path depths $\nabla W'(x,y)$ at predetermined points; simulated centroid positions similar to the aLIGO intermediate test masses.
    \item Add realistic Gaussian noise to the gradient fields, simulating centroiding algorithm noise.
    \item Randomly choose zero to five vectors to be replaced with large vectors in random directions, simulating ``hot pixel effects'' (imperfections in the Hartmann sensors).
    \item Store the beam, point absorber and hot pixel positions as labels for training, testing, and validation.
\end{enumerate}

\subsubsection{Modelling thermal deformation}
We utilise the analytical Hello-Vinet~\citep{Vinet2009} steady-state solutions for the thermal deformation of the test masses to generate the component of $W'(x,y)$ from uniform absorption. 
This simulates the thermalisation of the test masses for randomly chosen beam spot positions and laser powers. 
We utilise the \texttt{Finesse}\footnote{\url{https://zenodo.org/records/821364}} package \citep{Brown2020, Freise2004, Freise2013, Brown2014} implementation of this analytical model.
The component of $W'(x,y)$ from the point absorbers are then simulated by the model given in~\citep{Brooks2021} and added for each simulated point absorber.
These approximations of $W'(x,y)$ are considered adequate for the comparatively low spatial frequency sampling frequency of the Hartmann sensors.
The algorithm is intended to detect point absorbers and not for rigorous quantitative measurements of their severity.

\begin{figure*}
    \centering
    \includegraphics[width=0.9\linewidth]{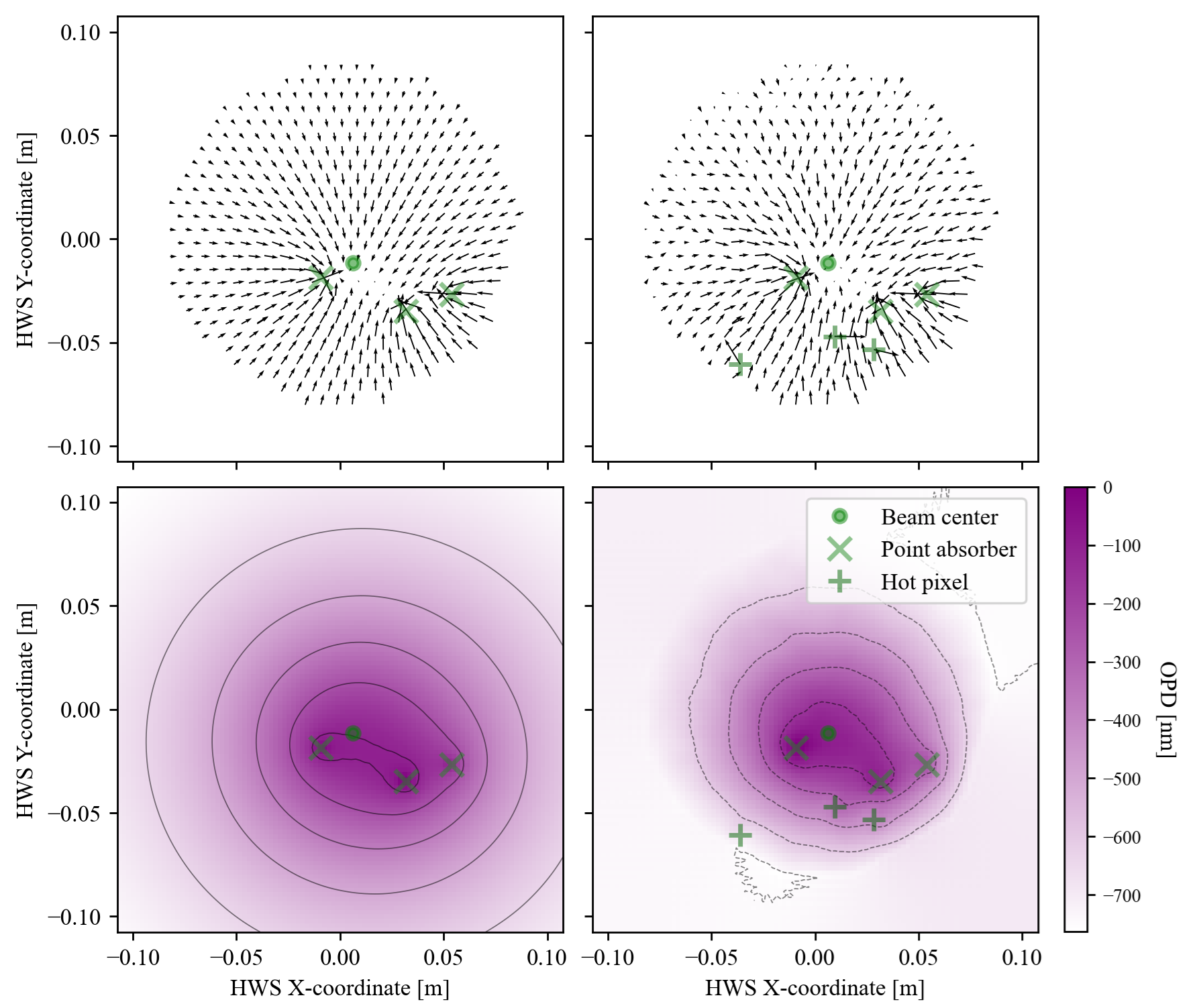}
    \caption{Simulated Hartmann wavefront sensor data. The noiseless gradient field (top left) is sampled from the Hello-Vinet optical path depth (bottom left). Gaussian scatter is added to each vector in the field to simulate centroiding errors (top right). Zero to five randomly selected vectors are also scattered by a large amount to simulate hot pixels. The noisy gradient field is visualised with optical path depth (bottom right), resembling aLIGO Hartmann wavefront sensor data.}
    \label{fig:sim-all}
\end{figure*}

We simulate point absorbers of varying severity and positions. 
We chose the number of point absorbers from a uniform distribution for each simulation on the interval (0, 5).
Each point absorber absorbs a fraction of the total power, which we chose from a uniform distribution on the interval (5\%, 10\%).
We chose the position ($r, \theta$) of the beam spot and each point absorber from uniform distributions.
The radius $r$ is defined on the interval $\unit[0]{cm}$--$\unit[2.5]{cm}$ for beam spots and $\unit[0]{cm}$--$\unit[6]{cm}$ for point absorbers.
The angle $\theta$ is defined on the interval $\unit[0]{rad}$--$\unit[2\pi]{rad}$.
Beam spot positions are relative to the centre of the test mass, and point absorber positions are relative to the beam spot position.
The output of this simulation is the noiseless theoretical optical path depth for a test mass with random point absorbers at thermal equilibrium.
Figure~\ref{fig:sim-all} (bottom left) shows an example of a noiseless optical path depth with three point absorbers surrounding a central heating beam.

\subsubsection{Simulating Hartmann wavefront sensor measurements}
These simulated optical path depths, $W'(x,y)$, are converted to simulated measurements of $\nabla W'(x,y)$ with a Hartmann wavefront sensor.
This is done by numerically calculating the gradient and sampling it at discrete gradients across the wavefront.
The positions at which the discrete gradients are taken have been generated to appear similar to the aLIGO ITMs.
For each simulation, we create a gradient field made of hundreds of vectors.
Figure~\ref{fig:sim-all} (top left) shows the sampled gradient field of the simulated optical path depth (bottom left).

\subsubsection{Simulating measurement noise and hot pixels}
Noise affects all vectors in the gradient field, causing varying degrees of distortion to their magnitude and direction.
Sources of noise in the gradient field arise from the Hartmann centroid detection algorithm, which measures how each centroid on the CCD moves between frames.
Fluctuations in the centroid detection algorithm affect each vector in the gradient field, causing minor Gaussian distortions. 
Electronic faults in the CCD, such as hot pixels, can cause the Hartmann centroid detection algorithms to fail, leading to large, erroneous vectors.
Low light intensity at the edges of the CCD or clipping of the probe beam can similarly cause issues with the centroid detection algorithm, leading to noisy edge vectors.

Gaussian noise is added to each vector to simulate the shot noise from the measurement process.
Two random numbers from a Gaussian distribution with mean-zero and width \unit[$5\times10^{-7}$]{rad} were chosen and added to each vector's $x$ and $y$ magnitude.
Failures in the centroiding algorithm were simulated by replacing a small number of vectors with large vectors in random orientations.
In each sample, the number of failed vectors is randomly chosen from a uniform distribution on the interval (0, 5).
The radial length of each failed vector is uniformly chosen on the interval of (2$\mu_v$, 4$\mu_v$), where $\mu_v$ is the mean length of vectors in the noisy gradient field.
The direction of each failed vector is uniformly chosen on the interval of (0, 2$\pi$).
Figure~\ref{fig:sim-all} (top right) shows these noise effects being applied to the noiseless gradient field (top left), including the addition of three hot pixels. 
The bottom right panel of Fig.~\ref{fig:sim-all} visualises the final product of our data generation method for comparison with the original, simulated optical path depth (bottom left).

\subsection{Training}\label{sec:methods-training}
Training was performed on a synthetic dataset of 10,000 gradient fields. 
Each point absorber in the dataset is labelled with a bounding box centred on the point absorber. Point absorber bounding boxes have a width and height of 7.5\% the size of the entire image ($\unit[20]{px}$, or $\sim\unit[1.5]{cm}$ in real space). This box size ensures that only nearby vectors are considered when predicting point absorbers. Each sample also contains a single beam spot, randomly positioned in a $\unit[2.5]{cm}$ radius of the test mass centre. A bounding box centred on the beam with doubled dimensions ($\unit[39]{px}$, or $\sim\unit[3]{cm}$) is added as a secondary predictive target.

The synthetic dataset was split into ratios of (80\%, 10\%, and 10\%) for training, validation, and testing, respectively. 
The model is trained for 50 epochs with a batch size of 8. During training, samples are ``augmented'' with various effects to boost model diversity and robustness. 
Here, ``augmentation'' refers to intentionally altering training data to increase data diversity and avoid biases during training.
Data augmentation, particularly in image-based machine learning, increases model robustness by diversifying patterns in training data, leading to better generalisations in pattern recognition.
We include data augmentation for this task to create a well-generalised machine-learning model.

Augmentations are applied randomly and with varying strengths from a uniform distribution between 0 and a maximum fractional value. 
Our model's augmentations included horizontal and vertical translation, rescaling, horizontal flipping, erasing and mosaic combination. Table~\ref{tab:augmentations} (in Appendix~\ref{sec:app-model}) provides descriptions of each augmentation and the chosen maximum fractional value.

After 50 epochs of training, training and validation losses continue to trend downwards, suggesting no signs of overfitting. 
Evaluation of the test set indicates the true-positive rate for point absorber detection is $>$0.99.
This gives us confidence that the training is completed with an appropriate number of epochs to produce a high-performance model.
More details on the training and evaluation of the machine-learning model are provided in Appendix~\ref{sec:app-model}.

The machine-learning model scans an input image and places boxes around objects it regards as point absorbers or central heating beams.
Each object the model identifies has an associated confidence value.
These confidence values should be taken as a relative ranking system so that, e.g., a point absorber candidate with a confidence of 0.9 is more confidently identified than one with a confidence of 0.5.
They do not represent $p$-values; for example, a candidate with a confidence of 0.7 will not necessarily be a false positive exactly 30\% of the time.

\section{Characterising the LIGO test masses}\label{sec:results}
\subsection{Comparison to aLOGs}
To verify that our model performs well on real data, we compare its predictions with human assessments of archival aLIGO data.
We analyse $\unit[10,000]{s}$ of data in each case, starting from laser power-up.
This duration allows us to study the appearance of point absorbers as the test mass transitions from the cold state to the hot state.
We produce videos with a frame rate of $\unit[30]{Hz}$ showing the Hartman gradient field with \texttt{YOLO} identification boxes overlaid.
Each video frame depicts a $\unit[20]{s}$ averaged gradient field of the test mass.
These videos are available on Zenodo\footnote{\url{https://zenodo.org/records/13958710}}.

\begin{figure*}
    \centering
    \includegraphics[width=0.8\linewidth]{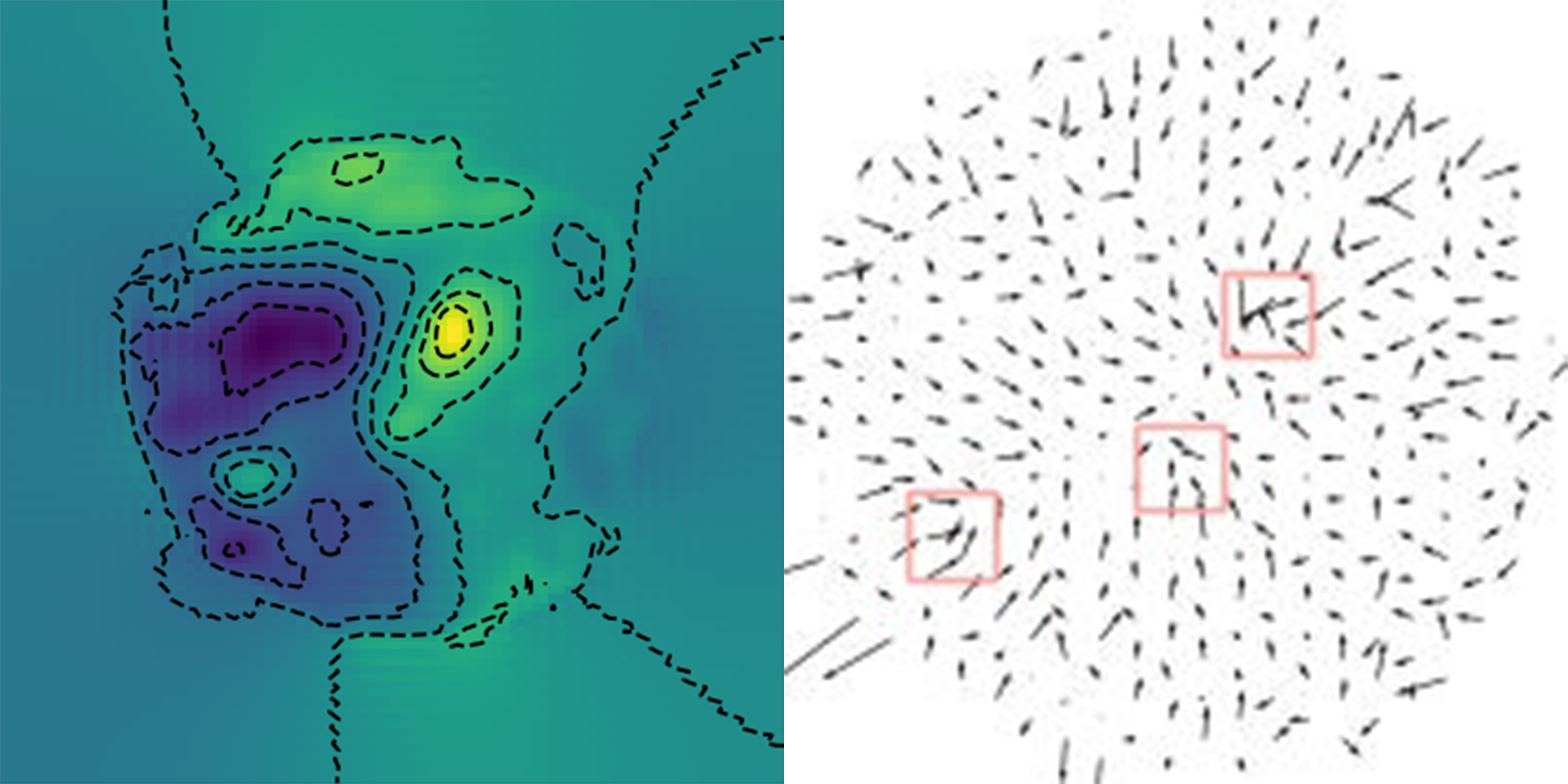}
    \caption{Side-by-side comparison of optical path depth (left) and \texttt{YOLO} predictions (right) for LLO alog \#60102.
    The optical path depth shows non-uniform heating of the Livingston ETMX, with clear excess absorption highlighted by the peaks.
    Experts concluded three point absorbers are responsible for these peaks.
    \texttt{YOLO} consistently classified these peaks as three point absorbers, matching the expert analysis with high confidence.
    }
    \label{fig:alog-60102}
\end{figure*}

The first case study is from the LIGO Livingston Observatory (LLO) alog \#60102, posted 19 May, 2022.\footnote{\url{https://alog.ligo-la.caltech.edu/aLOG/index.php?callRep=60102}}.
Figure~\ref{fig:alog-60102} (left panel) shows the optical path depth taken at an early phase of the power-up sequence. 
Using the optical path depth, the expert scientists who contributed to this alog identified three point absorber candidates, evidenced by two peaks near the centre of the test mass and one towards the left edge. 
The \texttt{YOLO} video (single frame shown in right panel) finds the same three point absorbers with high confidence values ranging from 0.8-0.9. 

\begin{figure*}
    \centering
    \includegraphics[width=0.7\linewidth]{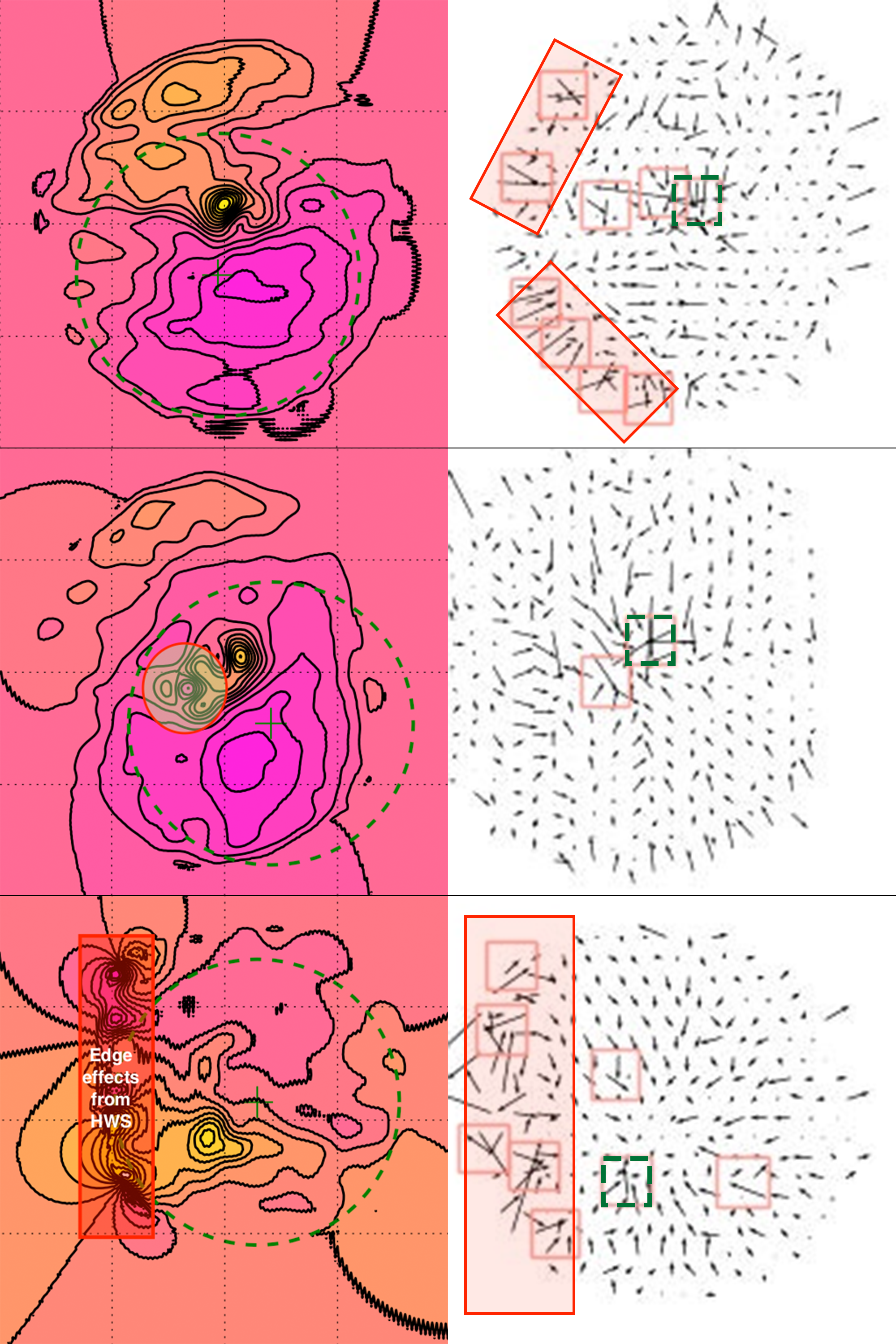}
    \caption{Side-by-side comparisons of optical path depth (left) and \texttt{YOLO} predictions (right) for LLO alog \#54588. 
    The three epochs evaluated in this alog are during O3a (top), O3b (middle) and post-O3b (bottom).
    Experts identified the appearance of a point absorber during O3b (circled in red), which was removed post-O3b.
    \texttt{YOLO} matches the experts' results, identifying the transient point absorber solely during the O3b epoch.
    \texttt{YOLO} also consistently identifies a known point absorber (green dashed boxes) across all epochs and identifies two previously unknown ones in the post-O3b epoch.
 \texttt{YOLO} detected many false positives along the edges of the O3a and post-O3b epoch (highlighted in red rectangles), which are easily identified by eye.
 \texttt{YOLO} also detected two false positives in the O3a epoch, where a vector from a genuine point absorber extended far towards the left side of the image and confused the algorithm.
    }
    \label{fig:alog-54588}
\end{figure*}

We next consider LLO alog \#54588, posted 22 March, 2022.\footnote{\url{https://alog.ligo-la.caltech.edu/aLOG/index.php?callRep=54588}} 
Figure~\ref{fig:alog-54588} shows three optical path depths from three epochs of the third observing run of LIGO (O3). From top-to-bottom, the epochs are during O3a, during O3b, and post O3b. 
LIGO scientists noted that a point absorber (highlighted in red) appeared during O3b and was subsequently removed after O3b.

Our algorithm produces consistent results, detecting the most prominent point absorber in all three epochs and the transient point absorber only during the O3b epoch. 
However, our analysis finds two additional point absorbers in the post-O3b epoch, highlighted in the top-left and top-right quadrants of Fig~\ref{fig:alog-54588}. 
One point absorber appears with a confidence of 0.8-0.85, while the other appears with a confidence of 0.5-0.6.
Our visual inspection of the associated Hartmann data convinces us that these candidate point absorbers are genuine.

Our model also classifies regions of the gradient field as point absorbers, which we believe to be false positives.
From this case, we identify two patterns that lead to false-positive detections.
The first stems from noisy vectors on the edges of the gradient field caused by low light intensity on the detector or harsh clipping of the probe beam.
This causes large vectors to appear randomly, resulting in sporadic point absorber-like patterns.
The inconsistency of these detections over time and their proximity to the edge helps to identify these false positives.
The second false-positive pattern can be seen in the O3a epoch, where a large vector causes nearby point absorber-like patterns.
These detections appear along the large vector and are consistent.
However, they can be ruled out through manual observation.

\begin{figure*}
    \centering
    \includegraphics[width=0.8\linewidth]{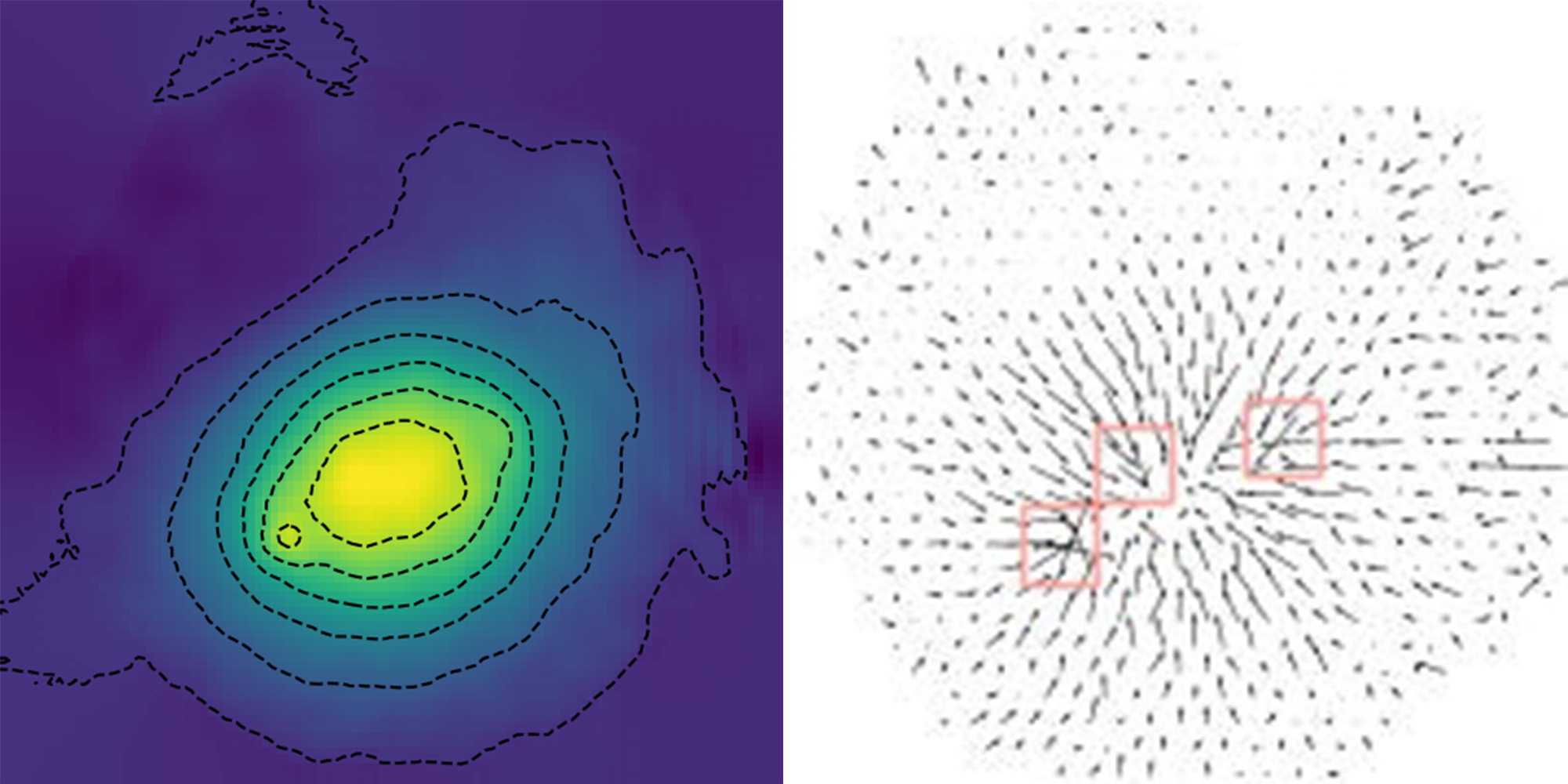}
    \caption{Side-by-side comparison of optical path depth (left) and \texttt{YOLO} predictions (right) for LLO alog \#72660.
    The optical path depth shows a central heating profile, with excess absorption of a point absorber highlighted by the contours.
    Experts identified one point absorber towards the bottom-left side of the beam, which coincided with the excess absorption.
    \texttt{YOLO} consistently classified this point absorber with high confidence and two additional, unidentified point absorbers.
    \texttt{YOLO} identified a point absorber to the right of the beam, coincident with a bulge in the contours, and another near the beam itself.
    }
    \label{fig:alog-72660}
\end{figure*}

The final case we consider is LLO alog \#72660, a recent beam spot movement test from 15 August, 2024.\footnote{\url{https://alog.ligo-la.caltech.edu/aLOG/index.php?callRep=72660}}
A point absorber was discovered on the Livingston ITMX during this beam spot test. 
Figure~\ref{fig:alog-72660} (left panel) shows the optical path depth of the ITMX.
There is a clear peak towards the bottom-left quadrant of the thermal lens.
Our model consistently detects this excess as a point absorber with a confidence score of $>$0.9.
Our model detects two additional point absorbers located near the beam spot and towards the right side of the beam spot.
These detections have confidence scores of 0.75 and 0.9, respectively. 
The high-confidence detection to the right of the beam correlates with a bulge in the optical path depth contours, indicative of a point absorber. 
The lower-confidence detection is more subtle due to its proximity to the beam spot. 
We believe this to be a previously unidentified point absorber.

\subsection{Ring heater tests}
We analyse data in which no known point absorbers are present to assess the algorithm's potential to produce false positives. 
We obtain Hartmann data during ring heater tests, in which the gradient field is expected to represent an inverted thermal lens.
Ring heaters~\cite{Lawrence2003,Brooks2016} counteract the main interferometer beam's thermal lensing profile by applying heat to the test mass edges.
These ring heater tests are conducted while the main interferometer beam is off.
With no heating beam, any point absorbers present do not absorb any power or appear in the optical path depth.
Therefore, any detections made by our model on ring heater tests are false positives.

\begin{figure*}
    \centering
    \includegraphics[width=0.8\linewidth]{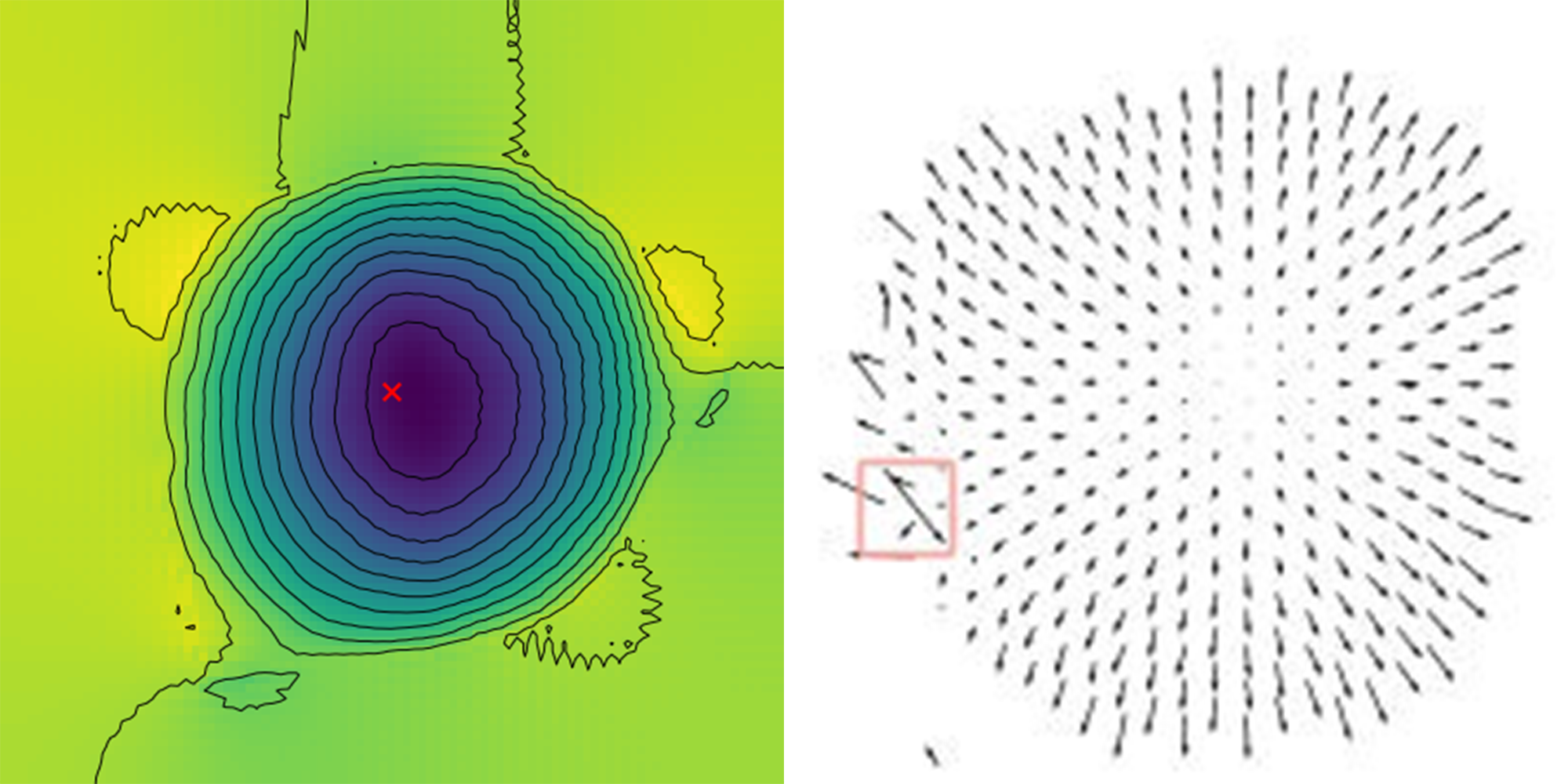}
    \includegraphics[width=0.8\linewidth]{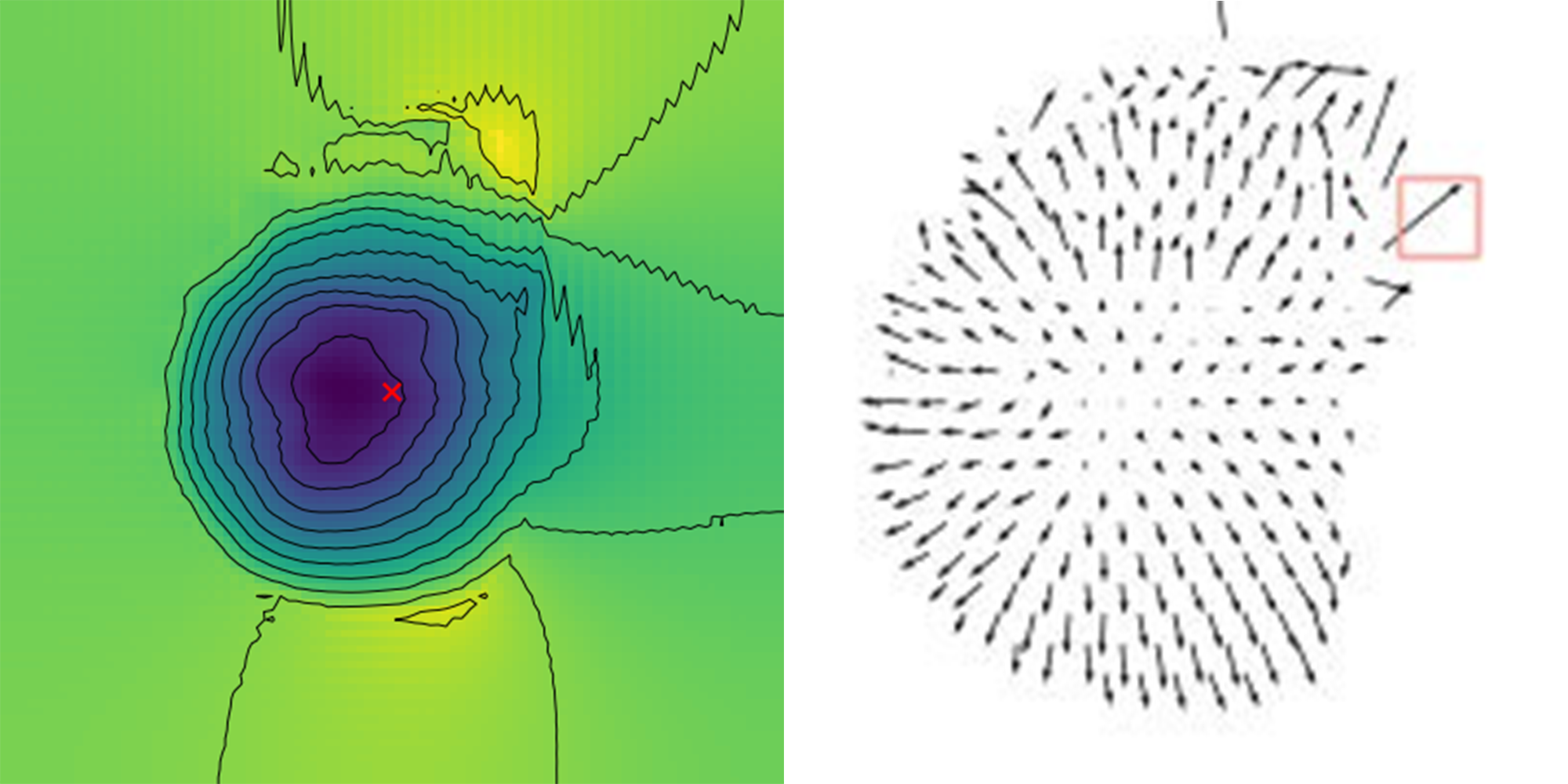}
    \caption{
    Side-by-side comparison of optical path depth (left) and \texttt{YOLO} predictions (right) for LLO alog \#69627.
    Ring heater tests are performed without the interferometer beam, so no point absorbers will appear in this alog.
    We use these tests to identify false-positive signals that cause \texttt{YOLO} to fail.
    The ITMX and ITMY produced no false positives, while the ETMX and ETMY each produced one.
    The ETMX (top) false positive is caused by a large misaligned vector, likely due to low light or a hot pixel on the edge of the Hartmann wavefront sensor.
    The ETMY (bottom) false positive is caused by the rare circumstance of a large vector pointing away from the center.
    This condition is not represented in the training set and, therefore, is a false positive caused by the training scope.
    }
    \label{fig:alog-69627}
\end{figure*}

Figure~\ref{fig:alog-69627} presents LLO alog \#69627, a series of ring heater tests conducted on 2 February, 2024 at the Livingston site.\footnotemark\space
We analyze the data for each test mass during the six-hour duration of the ring-heater test. 
The analysis of each test mass shows a gradient field diverging from the mirror centre, opposing thermal lensing. 
The ITMX and ITMY show no false positive point absorber detections during the test. 
The ETMX and ETMY independently show a false positive caused by anomalous vectors on the field's edge.
\footnotetext[7]{\url{https://alog.ligo-la.caltech.edu/aLOG/index.php?callRep=69627}}

The ETMX contains a single vector that points at an obtuse angle relative to the divergent field. 
This vector is likely caused by either an error in the centroiding algorithm in low light or a hot pixel. 
The vector causes the appearance of a small, point absorber-like convergence, leading to a false-positive.
The ETMY contains a single vector in line with the divergent gradient field.
Our algorithm consistently detects the tip of this vector as a false positive.
This false positive is likely an artifact of the training scope; the algorithm is operating on data it was not trained on.
Our algorithm has detected a large vector that points away from the mirror centre and is not near other vectors.
Since our algorithm was not trained on cases like this, it mistakenly classified the vector as a point absorber with confidence values between 0.5-0.75.

\section{Discussion/Conclusions and Future work}\label{sec:conclusions}

We have shown that machine learning can be an invaluable tool for efficiently identifying point absorbers.
Object detection algorithms can directly identify beam spots and point absorbers from the gradient field and reproduces the results of expert commissioners and even identifies previously undetected point absorbers.

We aim to deploy this algorithm within aLIGO to assist commissioners with fast, automated point absorber detection for constant monitoring of core optics.
Continued maintenance of our algorithm through retraining will yield more robust and higher-performing models. Training on transient thermal deformations and analysis of the temporal evolution of the gradient field may yield improved performances.
Although our algorithm was developed with aLIGO in mind, the principle is widely applicable. 
Other vacuum-isolated optical systems equipped with Hartmann wavefront sensors can utilise this algorithm to detect point absorbers on mirror surfaces.

We intend to further utilise \texttt{YOLO} for commissioning, with plans to investigate the short-term transient characteristics and long-term evolution of point absorbers.
In the future, next-generation gravitational-wave observatories can use this technology to prevent point absorbers from impacting operations and causing catastrophic damage to the instrument.
This work highlights the value that machine learning can bring to gravitational-wave observatories, instrumentation and commissioning efforts.
Further development of machine-learning algorithms is likely to profoundly impact these areas in the coming years.

\section{Acknowledgments}

The authors would like to thank the OzGrav ARC grants CE170100004 and CE230100016 that supported both running and attending the workshops that started this collaboration, and D. Brown for the support of ARC grant DE230101035. The authors would also like to acknowledge the support of the LIGO scientific collaboration (this manuscript has the DCC ID: LIGO-P2400533) computing services that were used to access and process the Hartmann images from the sites.

\appendix
\section{Additional details}\label{sec:app-model}

\begin{figure*}
    \centering
    \includegraphics[width=0.95\linewidth]{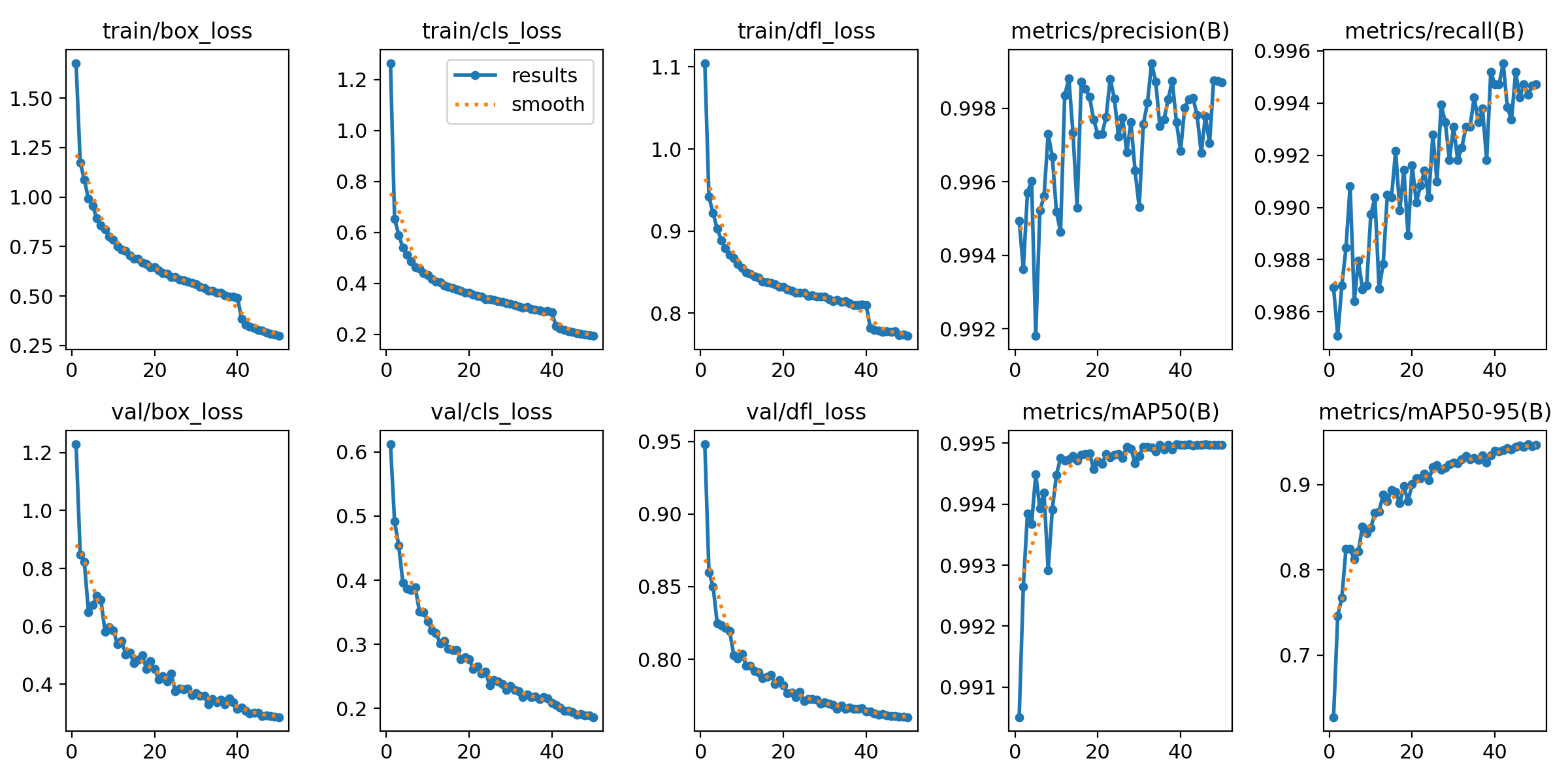}
    \caption{Losses and performance metrics of the \texttt{YOLO} algorithm during transfer learning. Validation and training losses from left to right are bounding box loss, classification loss and DFL loss \citep{Li2022}. Bounding box and DFL loss help the algorithm learn where to place bounding boxes and classification loss for what objects those boxes represent. Together, these losses train the model to detect specific objects in images. The steady decline and plateauing of these losses indicate the model is well-trained, with no under or overfitting.
    }
    \label{fig:yolo-losses}
\end{figure*}

\begin{figure*}
    \centering
    \includegraphics[width=.8\linewidth]{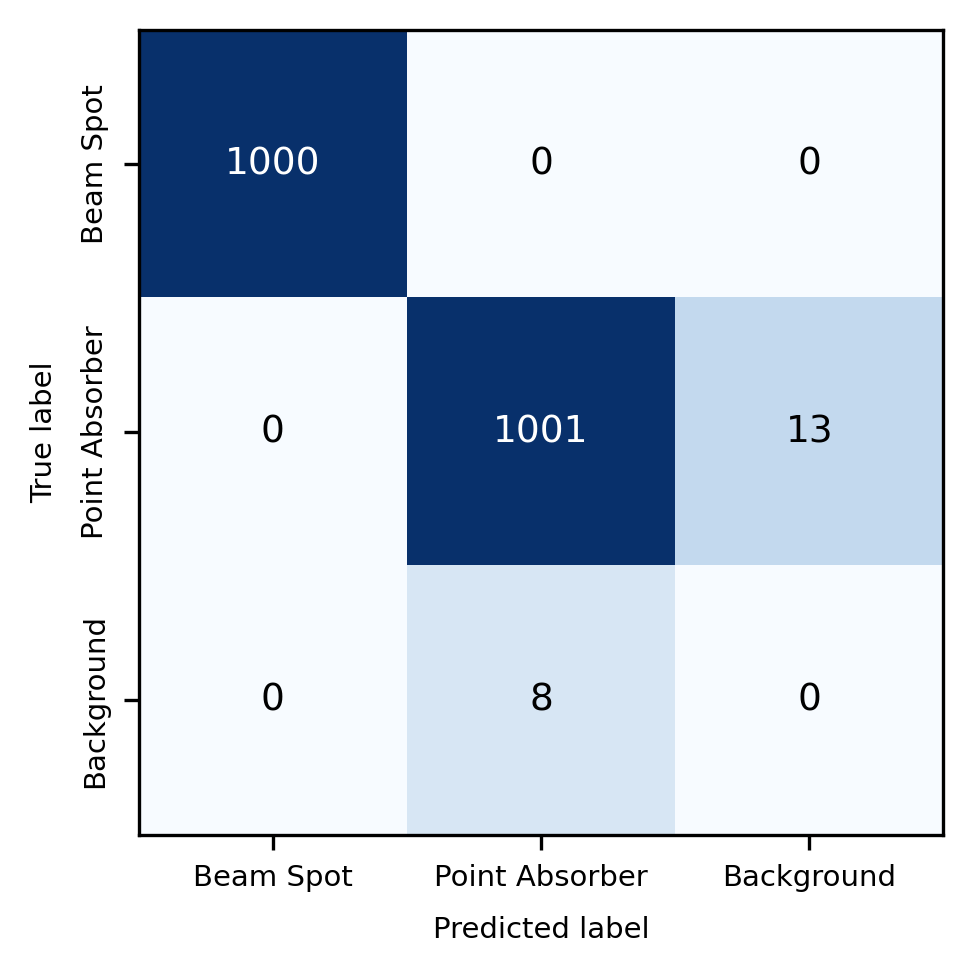}
    \caption{Confusion matrix of \texttt{YOLO} when applied to the test set.
    We show the transfer-learned model has a true-positive rate of $>0.99$, with rare false positives and false negatives for point-absorber detection.
    This provides confidence that our transfer-learned \texttt{YOLO} model will perform well with minimal classification errors.
    }
    \label{fig:yolo-CM}
\end{figure*}

Here, we present additional details relevant to training the \texttt{YOLO} algorithm. Figure \ref{fig:yolo-losses} shows the training and validation loss metrics\footnote{\url{https://github.com/ultralytics/ultralytics/blob/main/ultralytics/utils/loss.py}} as a function of the training epoch. We note the smoothly declining and plateauing losses indicate well-chosen training hyperparameters. These losses show no signs of underfitting or overfitting. Figure \ref{fig:yolo-CM} shows the confusion matrix of the designated test set. The confusion matrix shows our \texttt{YOLO} model achieves high precision and accuracy when classifying beam spots and point absorbers. Table \ref{tab:augmentations} describes the augmentations applied randomly to the data during training.

\begin{table*}[]
\resizebox{\textwidth}{!}{%
\begin{tabular}{lcr}
\hline
Augmentation    & Maximum Value & Description                                                                                                                                                                                                  \\ \hline
Translation     & 10\%          & \begin{tabular}[c]{@{}r@{}}Randomly shifts the image by a fraction of the image size independently in both axes. \\ Helps mitigate positional biases\end{tabular}                                            \\ \hline
Scale           & 50\%          & \begin{tabular}[c]{@{}r@{}}Randomly enlarges/shrinks the image by a fraction of the image size. \\ Introduces variability in image patterns; improves general pattern recognition.\end{tabular}              \\ \hline
Horizontal Flip & 50\%          & \begin{tabular}[c]{@{}r@{}}Randomly flips the image about a centered vertical axis. \\ Left-most pixels become right-most pixels and vice versa. \\ Improves variation within the training set.\end{tabular} \\ \hline
Erasing         & 40\%          & \begin{tabular}[c]{@{}r@{}}Randomly erases sections of the image. \\ Promotes deeper pattern recognition to detect partial objects.\end{tabular}                                                             \\ \hline
Mosaic          & 100\%         & \begin{tabular}[c]{@{}r@{}}Randomly creates a 2x2 composite image, combining four training samples into one. \\ Adds variability to the number of objects in training samples.\end{tabular}                  \\ \hline
\end{tabular}%
}
\caption{Augmentations applied to data during model training. Augmentations improve model robustness by increasing the variation within training set samples. Augmentations are applied randomly with a strength sampled uniformly between 0 and a maximum fractional value.}
\label{tab:augmentations}
\end{table*}

\bibliography{refs}
\end{document}